\documentclass[prl,twocolumn,showpacs,amsmath,amssymb,epsfig]{revtex4-1} 
 
\usepackage{graphicx}

\begin{document}

\title{Observation of Anomalous Internal Pair Creation in $^8$Be: \\ 
A Possible Signature of a Light, Neutral  Boson}

\author{A.J. Krasznahorkay}
\email{kraszna@atomki.hu}
\author{M. Csatl\'os}
\author{L. Csige}
\author{Z. G\'acsi}
\author{J. Guly\'as}
\author{M. Hunyadi}
\author{I.~Kuti}
\author{B.M. Nyak\'o}
\author{L. Stuhl}
\author{J. Tim\'ar}
\author{T.G. Tornyi}
\author{Zs. Vajta}
\affiliation{Institute for Nuclear Research, Hungarian Academy of
  Sciences (MTA Atomki), P.O. Box 51, H-4001 Debrecen, Hungary} 
\author{T.J. Ketel} \affiliation{Nikhef National Institute for Subatomic Physics, Science Park 105, 1098 XG Amsterdam, The Netherlands}
\author{A. Krasznahorkay} \affiliation{CERN, Geneva, Switzerland}
\affiliation{Institute for Nuclear Research, Hungarian Academy of
Sciences (MTA Atomki), P.O. Box 51, H-4001 Debrecen, Hungary}

\begin{abstract}  
Electron-positron angular correlations were measured for the {\it
  isovector magnetic dipole} 17.6 MeV state ($J^\pi=1^+$, $T=1$)
$\rightarrow$ ground state ($J^\pi=0^+$, $T=0$) and the {\it isoscalar
  magnetic dipole} 18.15 MeV ($J^\pi=1^+$, $T=0$) state $\rightarrow$
ground state transitions in $^{8}$Be.  Significant deviation from the
internal pair creation was observed at large angles in the angular
correlation for the isoscalar transition with a confidence level of $>
5\sigma$. This observation might indicate that, in an intermediate
step, a neutral isoscalar particle with a mass of 16.70$\pm0.35 $
(stat)$\pm 0.5 $ (sys) MeV$/c^2$ and $J^\pi = 1^+$ was created.

\end{abstract}

\pacs{23.20.Ra, 23.20.En, 14.70.Pw}

\maketitle

Recently, several experimental anomalies were discussed
as possible signatures for a new light particle \cite{fr12}.
Some predictions suggest light neutral bosons in the 10 MeV - 10 GeV mass
range as dark matter candidates, which couple to electrons and
positrons \cite{fa04,po08,da12,ho12}, to explain the anomalies.  
A number of attempts were made to find such particles by using
data from running facilities
\cite{me11,ab11,le12,ec12,ar12,ba13,ad13,ba08} or reanalyzing data of
preceding experiments \cite{bj09,an12,bl11,gn12,gni12}.  Since no
evidence was found, limits were set on their mass and their coupling
strength to ordinary matter. In the near future, ongoing experiments
are expected to extend those limits to regions in mass and coupling
strength which are so far unexplored. All of them are designed to
exploit the radiative production of the so-called dark photons
($\gamma^\prime$) by a very intense electron or positron beam on a
high-Z target \cite{es11,hps,be13,fr10,wo12,gn13}.

In the present work we reinvestigated the anomaly observed previously
in the internal pair creation of an isovector (17.6 MeV) 
and an isoscalar (18.15 MeV) M1 transitions in $^{8}$Be  \cite{bo96, bo97, bo01,
ti04,vi08,kr13}. 

 The expected signature of
the new particle is a very characteristic angular correlation of the
e$^+$e$^-$ pairs from its decay \cite{sa86,sa88}. Quantum electrodynamics (QED)
predicts \cite{ro49,sc81} that the angular correlation between the
$e^+$ and $e^-$ emitted in the internal pair creation (IPC) drops
rapidly with the separation angle $\theta$. In striking contrast, when
the transition takes place by emission of a short-lived
($\tau<10^{-13}$ s) neutral particle decaying into an $e^{+}e^{-}$
pair, the angular correlation becomes sharply peaked at larger
angles. The correlation angle of the two-particle decay (180$^\circ$
in the center-of-mass system) is decreased according to the Lorentz
boost in the laboratory system.


To populate the 17.6, and 18.15 MeV 1$^+$ states in $^{8}$Be
selectively, we used the $^{7}$Li(p,$\gamma$)$^{8}$Be reaction at the
$E_p$=0.441, and 1.03 MeV resonances \cite{ti04}.  Angular correlation of the
produced $e^+e^-$ pairs were detected in the experiments performed at
the 5 MV Van de Graaff accelerator in Debrecen. Proton beams with
typical current of 1.0 $\mu$A impinged on 15 $\mu$g/cm$^2$ thick
LiF$_2$ and 300 $\mu$g/cm$^2$ thick LiO$_2$ targets evaporated on 10
$\mu$m Al backings.

The $e^+e^-$ pairs were detected by five plastic $\Delta E$--$E$
detector telescopes similar to those built by Stiebing and
co-workers \cite{st04}, but we used larger telescope detectors in
combination with position sensitive detectors to increase the
coincidence efficiency by about 3 orders of magnitude.  $\Delta E$
detectors of 38$\times$45$\times$1 mm$^3$ and the $E$ detectors of
78$\times$60$\times$70 mm$^3$ were placed perpendicularly to the beam
direction at azimuthal angles of 0$^\circ$, 60$^\circ$, 120$^\circ$,
180$^\circ$ and 270$^\circ$.  These angles were chosen to obtain a
homogeneous acceptance of the $e^+e^-$ pairs as a function of the
correlation angle. The positions of the hits were registered by
multiwire proportional counters (MWPC) \cite{charpak} placed in front
of the $\Delta E$ and $E$ detectors.

The target strip foil was perpendicular to the beam direction.  The
telescope detectors were placed around the vacuum chamber made of a
carbon fiber tube. A detailed description of the experimental setup is
published elsewhere \cite{gu15}.

$e^{+}e^{-}$ pairs of the 6.05 MeV transition in $^{16}$O, and of the
4.44 MeV and 15.11 MeV transitions in $^{12}$C excited in the
$^{11}$B($p$,$\gamma$)$^{12}$C reaction (E$_p$= 1.6 MeV) were used to
calibrate the telescopes. $\gamma$ rays were also detected for
monitoring. A $\epsilon_{rel}$=20\% HPGe detector (measured at 1.33
MeV relative to that of a standard 3''-diameter, 3''-long NaI(Tl)
scintillator) was used at 50 cm from the target to detect the 477.61
keV $\gamma$ ray in the $^{7}$Li(p,p$^\prime\gamma$)
reaction \cite{as02}, which has a very high cross section and could be
used to follow the Li content of the target as a function of time.

In order to check the effective thickness of the targets during the
long runs, the shape (width) of the high energy $\gamma$ rays was
measured by a 100\% HPGe detector. In the case of the broad ($\Gamma$=
138 keV) 18.15 MeV resonance, the energy of the detected $\gamma$ rays
is determined by the energy of the proton at the time of its capture
(taking into account the energy loss in the target), so the energy
distribution of the $\gamma$ rays reflects the energy distribution of
the protons.  The intrinsic resolution of the detector was less than
10 keV at 17.6 MeV and the line broadening caused by the target
thickness was about 100 keV allowing us a reliable monitoring.

\begin{figure}[htb]
    \begin{center}
        {\includegraphics[scale=0.4]{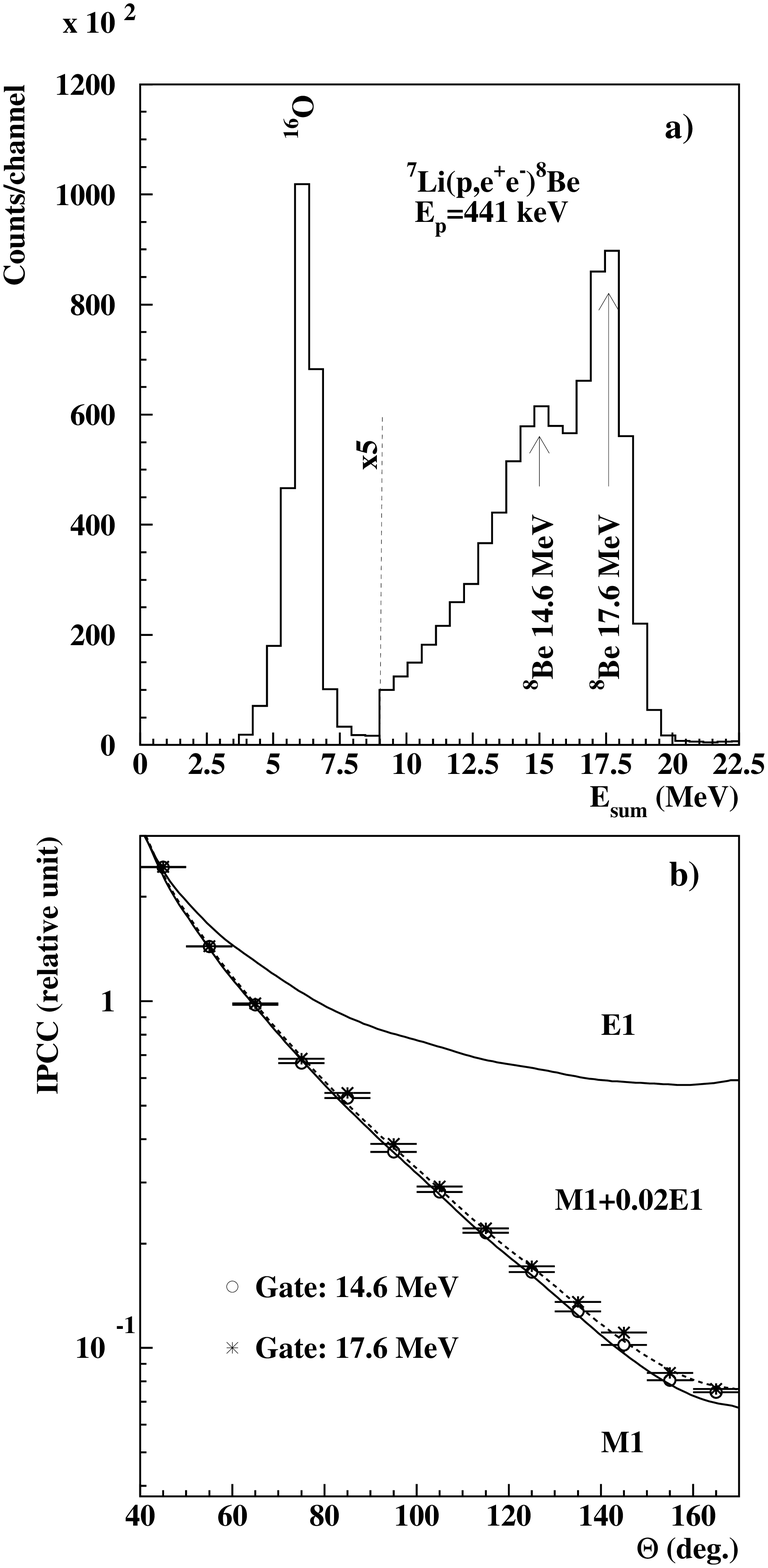}}\hspace{0.5cm}                      
\caption{\it Measured total energy spectrum (a) and angular correlation (b) of
  the $e^{+}e^{-}$ pairs originated from the decay of the 17.6 MeV
  resonance compared with the simulated angular correlations \cite{gu15} assuming
  M1 (full curve) and M1+2\%E1 mixed transitions (dashed line).}
\label{441}
    \end{center}
\end{figure}

Figure \ref{441} shows the total energy spectrum of $e^+e^-$ pairs
measured at the proton absorption resonance of 441 keV (a) and the angular
correlations  of the $e^{+}e^{-}$ pairs originated from the 17.6 MeV
$1^+ \rightarrow 0^+_1$ isovector M1 transition and the 14.6 MeV $1^+
\rightarrow 2^+_1$ transition (b).
 
The Monte Carlo (MC) simulations of the experiment were
performed using the GEANT code. Target chamber, target backing,
windows, detector geometries were included in the simulation in order
to model the detector response to $e^+e^-$ pairs and  $\gamma$
rays. The scattering of the $e^+e^-$ pairs and the effects of the
external pair creation in the surrounding materials could also be
investigated.
Beside the IPC process, the background of $\gamma$ radiation,
external pair creation (EPC) and multiple lepton scattering were
considered in the simulations
to facilitate a thorough
understanding of the spectrometer and the detector response \cite{gu15}.

We observed a slight deviation from the simulated internal pair
conversion correlation (IPCC) curve at large angles above 110$^\circ$
confirming the results of a previous measurement \cite{bo01}, but the
deviation  could be explained by admixing some E1 component from the
background. Previously, pure M1
transitions from the decay of the 17.6 MeV resonance were assumed
\cite{bo96,bo97,bo01}. It is true for the resonances itself, but not
for the underlying background, which is reasonably small (but not
negligible) for the 17.6 MeV resonance. The background originates from
the direct (non-resonant) proton capture and its multipolarity is
dominantly E1 \cite{dipole},  and it adds to the M1 decay of the
resonance.  The contribution of the direct capture depends on the
target thickness if the energy loss of the beam in the target is
larger than the width of the resonance.

As shown in Fig.\ref{441}b, the slope of the E1 angular correlation is
much smaller than the slope of the M1 one, so by adding even a small
contribution of E1 radiation, the angular correlation at large angles
is modified considerably. The dashed simulated curve in Fig. \ref{441}b
is obtained by fitting a small (1.4\%) E1 contribution to
the dominant M1 one, which describes the experimental data reasonably
well.

The 18.15 MeV resonance is much broader, $\Gamma$= 138 keV \cite{ti04}, 
than the one
at 17.6 MeV, $\Gamma$= 10.7 keV and its strength is more
distributed. The E1 contribution is expected to be larger than that of the
17.6 MeV resonance and, indeed, the deviation observed previously was much
bigger in the 75$^\circ$ -  130$^\circ$ angular region \cite{bo01}.
In the present work we extended the angular range to 170$^\circ$ and improved
the statistics
to check if the previously observed deviation can be explained with some E1
mixing also in this case.

\begin{figure}[htb]
    \begin{center}
        {\includegraphics[scale=0.4]{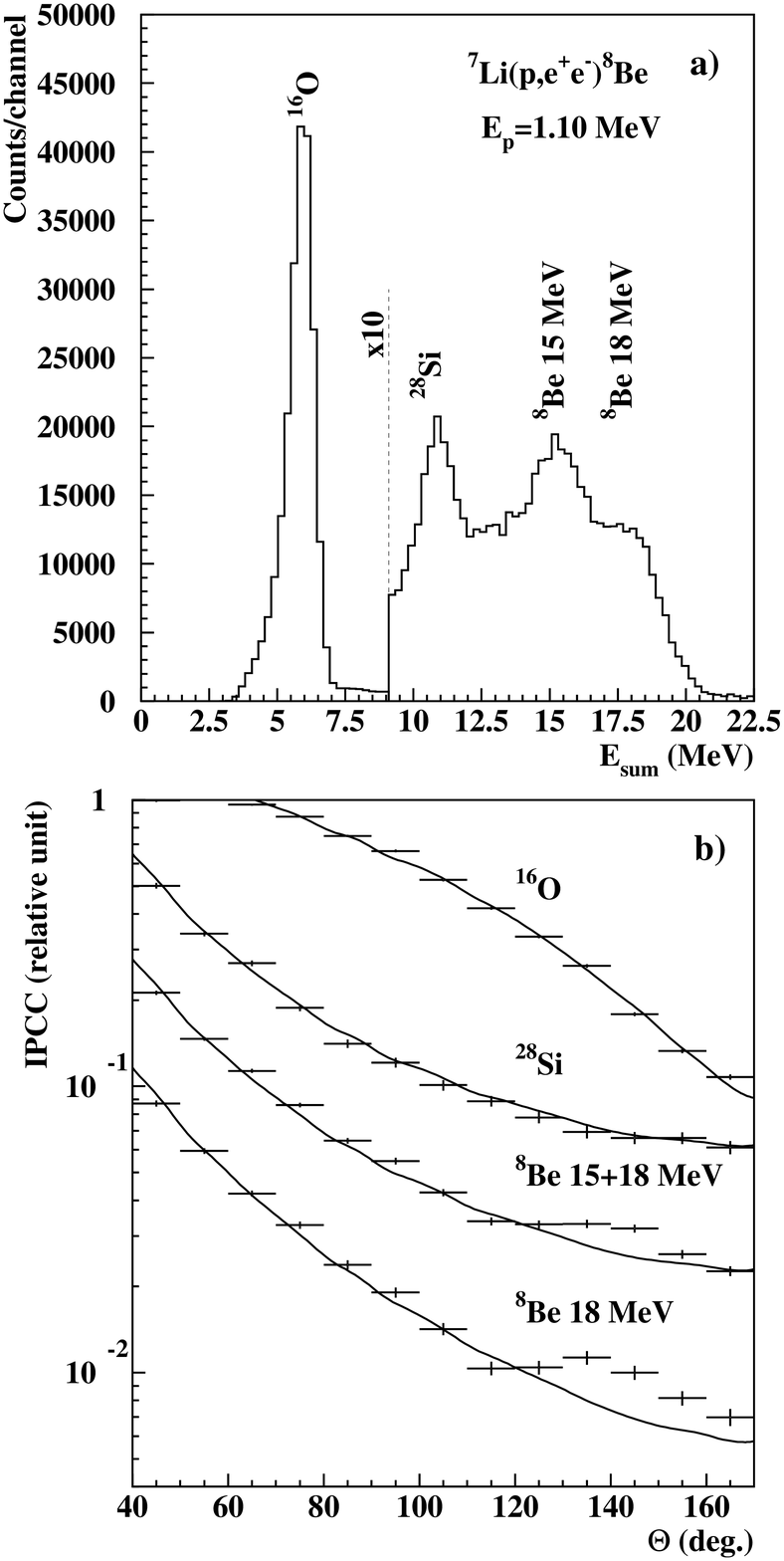}}\hspace{0.5cm}                      
\caption{\it Measured total energy spectrum (a) and angular
  correlations (b) of the $e^{+}e^{-}$ pairs created in the different
  transitions labelled in the figure, compared with the simulated
  angular correlations assuming E0 (from the $^{16}$O peak) and M1+E1
  mixed transitions from the other peaks.}
\label{1100}
    \end{center}
\end{figure}

Figure \ref{1100} shows the total energy spectrum of the $e^+e^-$
pairs measured at the proton absorption resonance of 1041 keV and the
angular correlation of the $e^{+}e^{-}$ pairs emitted in the 18 MeV
$1^+ \rightarrow 0^+_1$ isoscalar M1 transition and in the 15 MeV $1^+
\rightarrow 2^+_1$ transition. 

The spectra were obtained for symmetric
$-0.5\leq y\leq 0.5$ pairs, where the disparity (y) parameter is defined
as: 
$$y=(E_{e^-}-E_{e^+})/(E_{e^-}+E_{e^+}) \ , $$ 
where $E_{e^-}$ and
$E_{e^+}$ denote the kinetic  energies of the electron and positron,
respectively.

The acceptance as a function of the correlation angle in comparison to
isotropic emission was determined from the same data-set by using uncorrelated
$e^+e^-$ pairs of different single electron events \cite{gu15}.  With
this experimental acceptance, the angular correlations of different IPC
lines in Fig. 2a were determined simultaneously.

The 6.05 MeV E0 transition in $^{16}$O is due to the
$^{19}$F(p,$\alpha$)$^{16}$O reaction on a target contamination. The
11 MeV peak contains M1 and E1 transitions in $^{28}$Si. As shown in
Fig. \ref{1100} both the $^{16}$O and the $^{28}$Si angular
correlations can be well explained by the simulations.

The angular correlation for M1 transitions in $^8$Be in the 15+18 MeV
region (wide gate) shows a clear deviation from the simulations.  If
we narrow the gate around 18 MeV the deviation in the angular
correlation at around 140 degrees is even larger, so the deviation can be
associated with the 18 MeV transition, and can not be
explained by any amount of E1 mixing.

\begin{figure}[htb]
    \begin{center}
        {\includegraphics[scale=0.4]{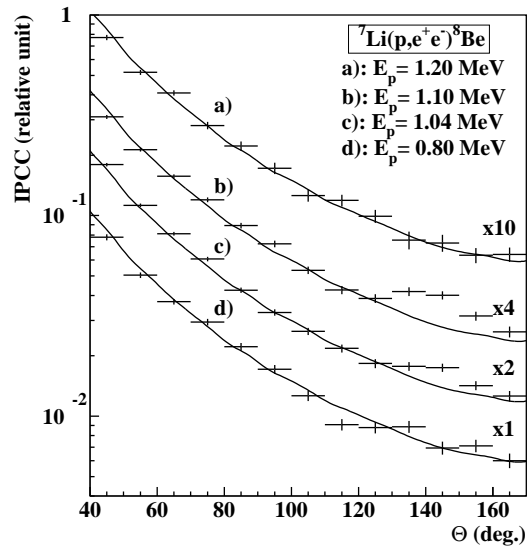}}\hspace{0.5cm}                      
\caption{\it Measured angular correlations of the $e^{+}e^{-}$ pairs
  originated from the ground state decay of the
  $^{7}$Li(p,$\gamma$)$^{8}$Be reaction (dots with error bars)
  compared with the simulated ones (full curves) assuming M1+E1 mixed
  transitions with the same mixing ratio for all curves at different
  beam energies.  }
\label{exp}
    \end{center}
\end{figure}

The angular correlation of the $e^+e^-$ pairs arising from the 
$\approx$18 MeV IPC transitions to the ground state 
excited in the $^{7}$Li(p,$\gamma$)$^{8}$Be reaction was
measured at different bombarding energies.  
The results are presented in Fig. \ref{exp}.

The pair correlation spectra measured at different bombarding energies are 
multiplied with different factors (indicated in the figure) for better 
separation.
 The full curves show
the IPC background (M1+23\%E1). 
 The deviation observed at the bombarding energy of
E$_p$=1.10 MeV (b) and at $\Theta\approx 140^\circ$ has a significance
of 6.8 standard deviations, corresponding to a background fluctuation
probability of $5.6\times10^{-12}$.
On resonance (b) the M1 contribution should be larger, so the background
should decrease faster than in other cases, which would make the deviation 
even larger and more significant.

\begin{figure}[htb]
    \begin{center}
        {\includegraphics[scale=0.4]{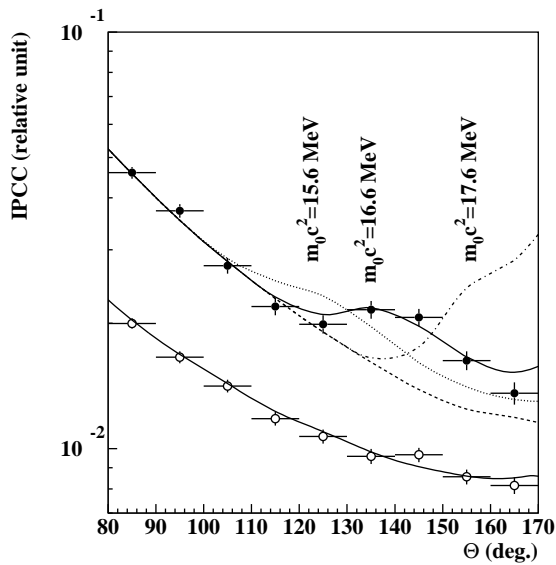}}\hspace{0.5cm}                      
\caption{\it 
Experimental angular $e^{+}e^{-}$ pair correlations measured in the
 $^7$Li(p,$e^{+}e^{-}$) reaction at 
E$_p$=1.10 MeV with -0.5 $\leq$ y $\leq$ 0.5 (closed circles) and $\vert y \vert\geq$ 0.5 
(open circles). The results of simulations of boson decay pairs 
added to those of IPC pairs are shown for different boson masses as 
described in the text.}
\label{boson-exp-simu}
    \end{center}
\end{figure}

The  $e^{+}e^{-}$ decay of a hypothetical boson emitted
isotropically from the target has been simulated together with the
normal IPC emission of $e^{+}e^{-}$ pairs.  
The sensitivity of the angular correlation 
measurements to the mass of the assumed boson is illustrated in 
Fig. \ref{boson-exp-simu}. 

Figure
\ref{boson-exp-simu} shows the experimental 
angular correlation of the $e^{+}e^{-}$ pairs in the narrow E$_{sum}$ = 18 MeV 
region and with -0.5 $\leq$ y $\leq$ 0.5 (full circles) together with the 
results of the simulations assuming
boson masses of $m_0c^2=$ 15.6 (dotted line), 16.6 (full curve) and
17.6 MeV (dash-dotted line), and the
simulation without assuming any boson contribution (dashed line).

Taking into account an IPC coefficient of $3.9\times 10^{-3}$ for the 18.15
MeV M1 transition \cite{ro49},
a boson to $\gamma$ branching
ratio of $5.8\times10^{-6}$ was was found for the best fit and was then used
for the other boson masses in Fig.\ref{boson-exp-simu}.

According to the simulations, the contribution of the assumed boson
should be negligible for asymmetric pairs with
 $0.5\leq\vert y \vert \leq 1.0$. The open circles with error bars in Fig.
\ref{boson-exp-simu} show the experimental data obtained for
asymmetric pairs (setting a wide, 15+18 MeV gate to get more statistics, as shown in Fig. 
 \ref{1100}b, and rescaled 
for better separation) compared with the simulations
(full curve) including only M1 and E1 contributions.

\begin{figure}[htb]
    \begin{center}
        {\includegraphics[scale=0.4]{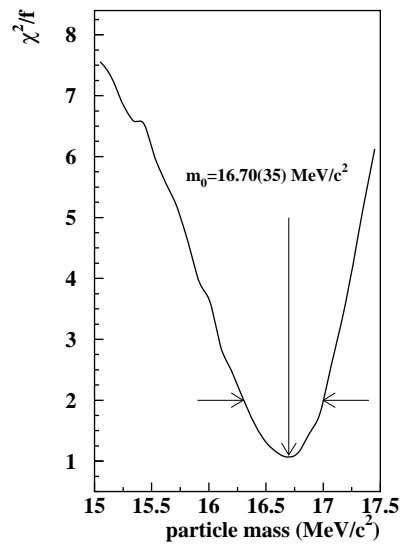}}\hspace{0.5cm}                      
\caption{\it Determination of the mass of the new particle by the $\chi^{2}/f$ 
method, by comparing the experimental data with the results of the simulations
obtained for different particle masses.
        }
\label{chi2}
    \end{center}
\end{figure}

The $\chi^2$ analysis mentioned above to judge the significance
of the observed anomaly was extended to extract the mass of the hypothetical
boson.
The simulated angular correlations included contributions from bosons
with masses between
 $m_0c^2=$ 15 and 17.5 MeV. The reduced $\chi^2$ values as a function of
the particle mass are shown in Fig. \ref{chi2}. 

As a result of the
$\chi^2$ analysis, we determined the boson mass to be $m_0c^2=$
16.70$\pm0.35 $ (stat) MeV. The minimum value for the $\chi^2$/f was 1.07.  
A systematic error caused by the
instability of the beam position on the target, as well as the
uncertainties in the calibration and positioning of the detectors is
estimated to be $\Delta\Theta=6^\circ$, which corresponds to 0.5 MeV
uncertainty in the boson mass.

In conclusion, we have measured the $e^+e^-$ angular correlation in
internal pair creation for the the M1 transitions depopulating the
17.6 and 18.15 MeV states in $^8$Be, and observed anomalous IPC in the
latter transition. The observed deviations from the M1 IPC in case of
the 17.6 MeV transition could be explained by the contribution of the
direct proton capture which presumably induce E1 transitions. However, E1
mixing alone cannot explain the measured anomaly in the 18 MeV pair
correlation.  The deviation between the experimental and theoretical
angular correlations is significant and can be described by assuming
the creation and subsequent decay of a boson with mass $m_0c^2$ =16.70$\pm0.35
$(stat)$\pm 0.5 $(sys) MeV$/c^2$ MeV.  The branching ratio of the
$e^+e^-$ decay of such a boson to the $\gamma$ decay of the 18.15 MeV
level of $^8$Be is found to be $5.8\times10^{-6}$ for the best fit.

Such a boson might be a good candidate for the relatively light
U(1)$_d$ gauge boson \cite{fa04}, or the light mediator of the
secluded WIMP dark matter scenario \cite{po08} or the dark Z (Z$_d$)
suggested for explaining the muon anomalous magnetic moment
\cite{ho12}. The coupling constant ($\epsilon^2$) of the dark Z having
a mass of 18 MeV is predicted to be in the 10$^{-6}$ range for
explaining the g-2 anomaly \cite{ho12}, which could fairly well 
explain the boson
to $\gamma$-decay branching ratio measured in the present work.
The lifetime of the boson with the above coupling strength is expected 
to be in the order of 10$^{-14}$ s \cite{da12}. This gives a flight distance 
of about 30 $\mu$m in the present experiment, and would imply a very sharp
resonance ($\Gamma\approx$ 0.07 eV) in the future  
$e^+e^-$  scattering experiments.

\section{Acknowledgements}
 
The authors are grateful to late Fokke de Boer for suggesting the above
studies and motivating us to improve the quality of the experimental
data. We regret that he could only contribute to early stages of this work.
We wish to thank Z. Tr\'ocs\'anyi and D. Horv\'ath for reading the manuscript
and for the many useful discussions.
 This work has been supported
by the Hungarian OTKA Foundation No.\, K106035, and by the European
Community FP7 - Contract ENSAR n$^\circ$ 262010.

\end{document}